%% file: hadron2011.tex
\documentclass[a4paper,11pt]{article}

\usepackage{contribution}

\input{econfmacros}
\input{contributionmacros}

\begin{document}

\input{contribution}

\end{document}

%% file: econfmacros.tex
\newcommand{\weblink}[2][]{%
    \ifthenelse{\equal{#1}{}}%
    {\textnormal{\url{#2}}}%
    {\textnormal{\href{#2}{#1}}}%
}

\newcommand{\acknowledgements}[1]{%
  \bigskip\bigskip
  \textsf{\textbf{\Large Acknowledgements}} \\[2ex]
  {#1}
  \bigskip
}

\def\beq{\begin{equation}}
\def\eeq#1{\label{#1}\end{equation}}
\def\eeqn{\end{equation}}

\def\beqa{\begin{eqnarray}}
\def\eeqa#1{\label{#1}\end{eqnarray}}
\def\eeqan{\end{eqnarray}}

\let\bar=\overbar

\def\Dslash{\not{\hbox{\kern-4pt $D$}}}
\def\dslash{\not{\hbox{\kern-2pt $\del$}}}

\def\msb{{\bar{\ssstyle M \kern -1pt S}}}

%% file: contributionmacros.tex
\newcommand{\contribution}[7][]{%
  \clearpage
  \thispagestyle{plain}
  \ifthenelse{\equal{#1}{}}
  {\hypersetup{pdftitle={#2}}}
  {\hypersetup{pdftitle={#1}}}
  \hypersetup{pdfauthor={{#3} {#4}}}
  {\centering\normalfont\LARGE\bfseries\sffamily #2 \par\nobreak}
  \lhead{}
  \chead{%
    \textit{\footnotesize XIV International Conference on Hadron Spectroscopy
      (\weblink[\textit{hadron2011}]{http://www.hadron2011.de}), 13-17 June 2011, Munich, Germany}%
  }
  \rhead{}
  \bigskip
  \begin{center}
    {#3} {#4}\ifthenelse{\equal{#6}{}}{}{\footnote{\weblink[#6]{mailto:#6}}}
    \ifthenelse{\equal{#7}{}}{}{#7} \\
    \textit{#5}
  \end{center}
  \bigskip
}

\renewcommand{\abstract}[1]{%
  \begin{center}
    \begin{minipage}{0.85\textwidth}
      \begin{footnotesize}
        #1
      \end{footnotesize}
    \end{minipage}
  \end{center}
  \bigskip
}

%% file: contribution.tex
%
%
%
%
%
{  

\makeatletter
\@ifundefined{c@affiliation}%
{\newcounter{affiliation}}{}%
\makeatother
\newcommand{\affiliation}[2][]{\setcounter{affiliation}{#2}%
  \ensuremath{{^{\alph{affiliation}}}\text{#1}}}
%

\contribution[$D$ and $D_s$ meson spectroscopy]  
{$D$ and $D_s$ meson spectroscopy from lattice QCD}  
{Daniel}{Mohler}  
{TRIUMF\\
  4004 Wesbrook Mall, Vancouver, BC \\
  V6T 2A3, CANADA}  
{mohler@triumf.ca}  
{\!\! and R.~M.~Woloshyn}
%

\abstract{%
We present results for the low-lying spectrum of $D$ and $D_s$ mesons from a lattice QCD calculation on 2+1 flavor Clover-Wilson configurations generated by the PACS-CS collaboration. In particular S- and P-wave states of charmed and charmed-strange mesons are explored for pion masses down to 156MeV. For the heavy quark, the Fermilab method is employed. In addition to ground states, some excited states are extracted using the variational method. To check our setup, calculations of the charmonium spectrum are also carried out. For charmonium, the low-lying spectrum agrees favorably with experiment. For heavy-strange and heavy-light systems substantial differences in comparison to experiment values remain in channels with nearby scattering states.
}
%

\section{Introduction}

The spectrum of charmed mesons contains states for which quark model expectations \cite{Godfrey:1985xj} did not hold. In particular, the charmed-strange mesons $D_{s0}^\star(2317)$ and $D_{s1}(2460)$, which in the limit of an infinitely heavy quark form a pair of mass-degenerate states with $j^P=\frac{1}{2}$(where $j$ the total angular momentum of the light quark and $P$ is parity) turn out to be narrow states with masses below the $DK$ and $D^\star K$ thresholds. Due to their unanticipated properties, it has been conjectured that these states are not of a simple $\bar{q}q$ nature. Lattice QCD is ideally suited to calculate the properties of hadrons from first principles. In recent years conceptional and algorithmic improvements have enabled calculations with light dynamical quarks in boxes of size $\approx$2.5fm and with fine lattice spacings. We present results from a recent calculation of charmonium and charmed mesons using dynamical gauge configurations with light sea quarks corresponding to pion masses as low as 156MeV, which is a significant improvement over previous simulations which employed either the quenched approximation or very heavy sea quarks. The following section summarizes the technical aspects of our simulation and the last section presents selected results. For more details please refer to the full published results \cite{Mohler:2011ke}.

\section{Technicalities}

For this study 2+1 Clover-Wilson configurations generated by the PACS-CS collaboration \cite{Aoki:2008sm} have been used. While the mass of the strange quark is kept close to its physical mass, pions made from light up and down quarks range from 702 MeV to 156 MeV. The number of lattice points is $32^3\times64$ and the lattice spacing has been determined \cite{Aoki:2008sm} to be $0.907(13)$fm. Table \ref{paratable} shows some of the run parameters and the number of configurations used.
\begin{table}[bht]
\begin{center}
\begin{tabular}{|c|c|c|c|c|}
\hline
 Ensemble & $c_{sw}^{(h)}$ & $\kappa_{u/d}$ &  $\kappa_s$  & \#configs $D/D_s$\\
\hline
 1 & 1.52617 & 0.13700 & 0.13640 & 200/200\\
 \hline
2 & 1.52493 & 0.13727 & 0.13640 & -/200\\
\hline
 3 & 1.52381 & 0.13754 & 0.13640 & 200/200\\
\hline
 4 & 1.52327 & 0.13754 & 0.13660 & -/200\\
\hline
 5 & 1.52326 & 0.13770 & 0.13640 & 200/348\\
\hline
 6 & 1.52264 & 0.13781 & 0.13640 & 198/198\\
\hline
\end{tabular}
\end{center}
\caption{\label{paratable}Run parameters for the PACS-CS lattices \cite{Aoki:2008sm}. All gauge configurations have been generated with the inverse gauge coupling $\beta=1.90$ and the light quark clover coefficient $c_{sw}^{(l)}=1.715$. The quantity $c_{sw}^{(h)}$ denotes the heavy quark clover coefficient.}
\end{table}

For the charm quarks we use the \emph{Fermilab method} \cite{ElKhadra:1996mp,Oktay:2008ex,Bernard:2010fr}. In this approach the spin-averaged \emph{kinetic mass} of the charmed-strange mesons is tuned to assume its physical value. Once the tuning has been done, differences in the rest masses are to be compared to experiment. The resulting charm quark hopping parameter $\kappa_c$ is listed in Table \ref{paratable}.

The spectrum results are obtained using the variational method \cite{Luscher:1990ck,Michael:1985ne}. For a given set of quantum numbers a matrix $C(t)_{ij}$ of interpolating fields is constructed
\begin{align}
C(t)_{ij}&=\sum_n\mathrm{e}^{-tE_n}\left <0|O_i|n\right>\left<n|O_j^\dagger|0 \right>.
\end{align}
On each time slice the generalized eigenvalue problem (GEVP) is solved
\begin{align}
C(t)\vec{\psi}^{(k)}&=\lambda^{(k)}(t)C(t_0)\vec{\psi}^{(k)} ,\\
\lambda^{(k)}(t)&\propto\mathrm{e}^{-tE_k}\left(1+\mathcal{O}\left(\mathrm{e}^{-t\Delta E_k}\right)\right).\nonumber
\end{align}
Asymptotically only a single state contributes to each eigenvalue. For details of our quark sources, which contain Jacobi-smeared \cite{Gusken:1989ad,Best:1997qp} and derivative sources \cite{Lacock:1996vy,Gattringer:2008be} please refer to the published results \cite{Mohler:2011ke}.

\section{Results}

In this section we present a selection of  results for charmed mesons and for charmonium. The charmonium spectrum below the $DD$ and $D^\star D$ thresholds is a good test case for our setup\footnote{in particular with regard to the discretization of the heavy charm quark}, as it contains only well established and uncontroversial states that are all believed to be predominantly of a $\bar{q}q$ nature and as our tuning procedure uses no input from the charmonium spectrum. The results we obtain for the lightest sea quarks are plotted in the left panel of Figure \ref{overviewplots}. Some additional results are also shown in Table \ref{charmtable}. In general our charmonium results agree qualitatively with the experimental spectrum. The Spin-dependent splittings displayed in Table \ref{charmtable} are expected to be sensitive to discretization effects and we expect that these are the dominant reason for underestimating the splittings compared to experiment.

\begin{table}[bht]
\begin{center}
\begin{tabular}{|c|c|c|}
\hline
 Mass difference & Our results [MeV] & Experiment [MeV]\\
\hline
\hline
1S hyperfine & $97.8\pm0.5\pm1.4$ & $116.6\pm1.2$\\
\hline
1P spin-orbit & $37.5\pm2.4\pm0.5$ & $46.6\pm0.1$\\
\hline
1P tensor & $10.44\pm1.13\pm0.15$ & $16.25\pm0.07$\\
\hline
2S hyperfine & $48\pm18\pm1$& $49\pm4$\\
\hline
\end{tabular}
\caption{\label{charmtable}Spin dependent mass splitting in the Charmonium spectrum.}
\end{center}
\end{table}

\begin{figure}[tb]
  \begin{center}
    \includegraphics[width=7.5cm,clip]{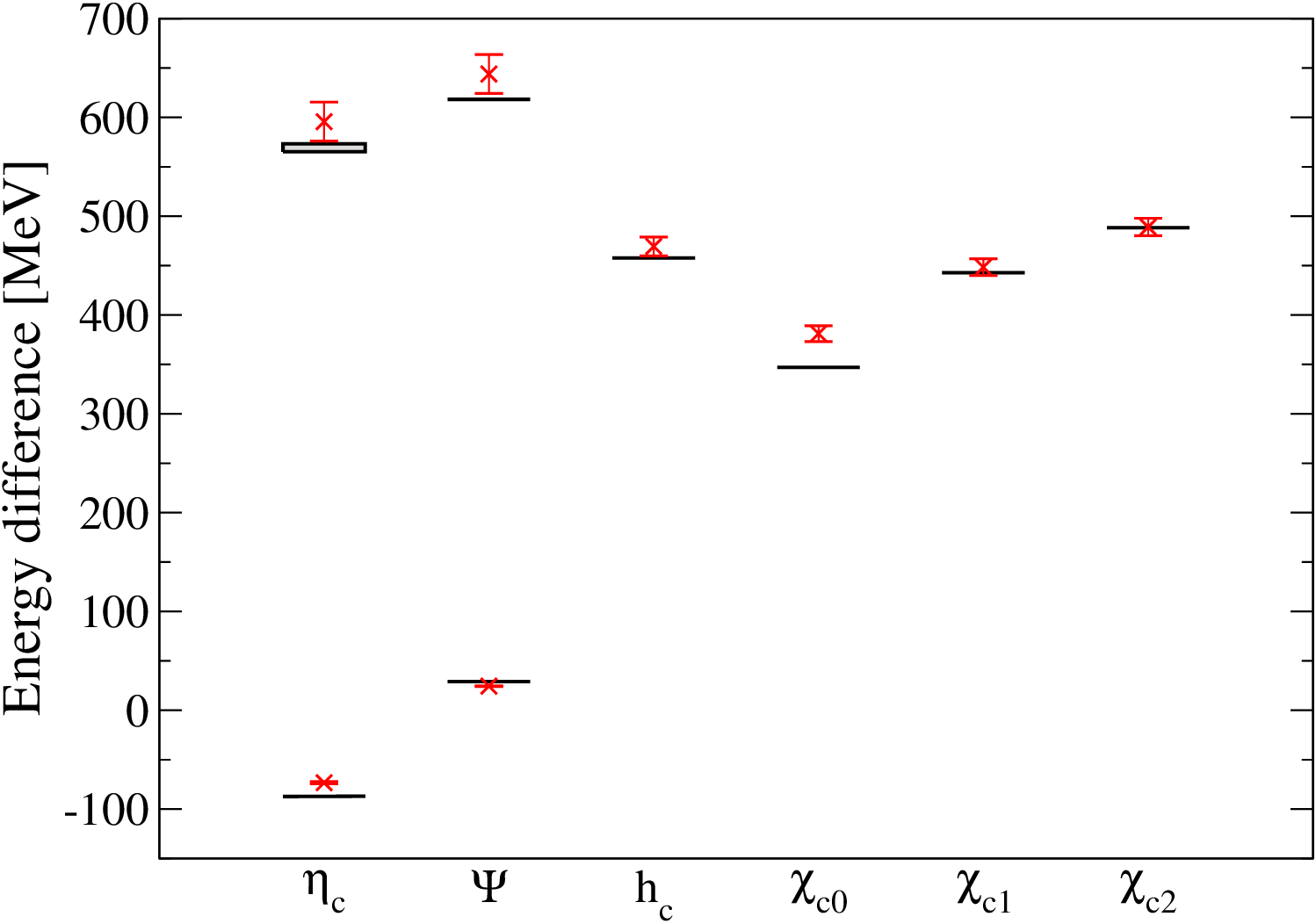}
    \includegraphics[width=7.5cm,clip]{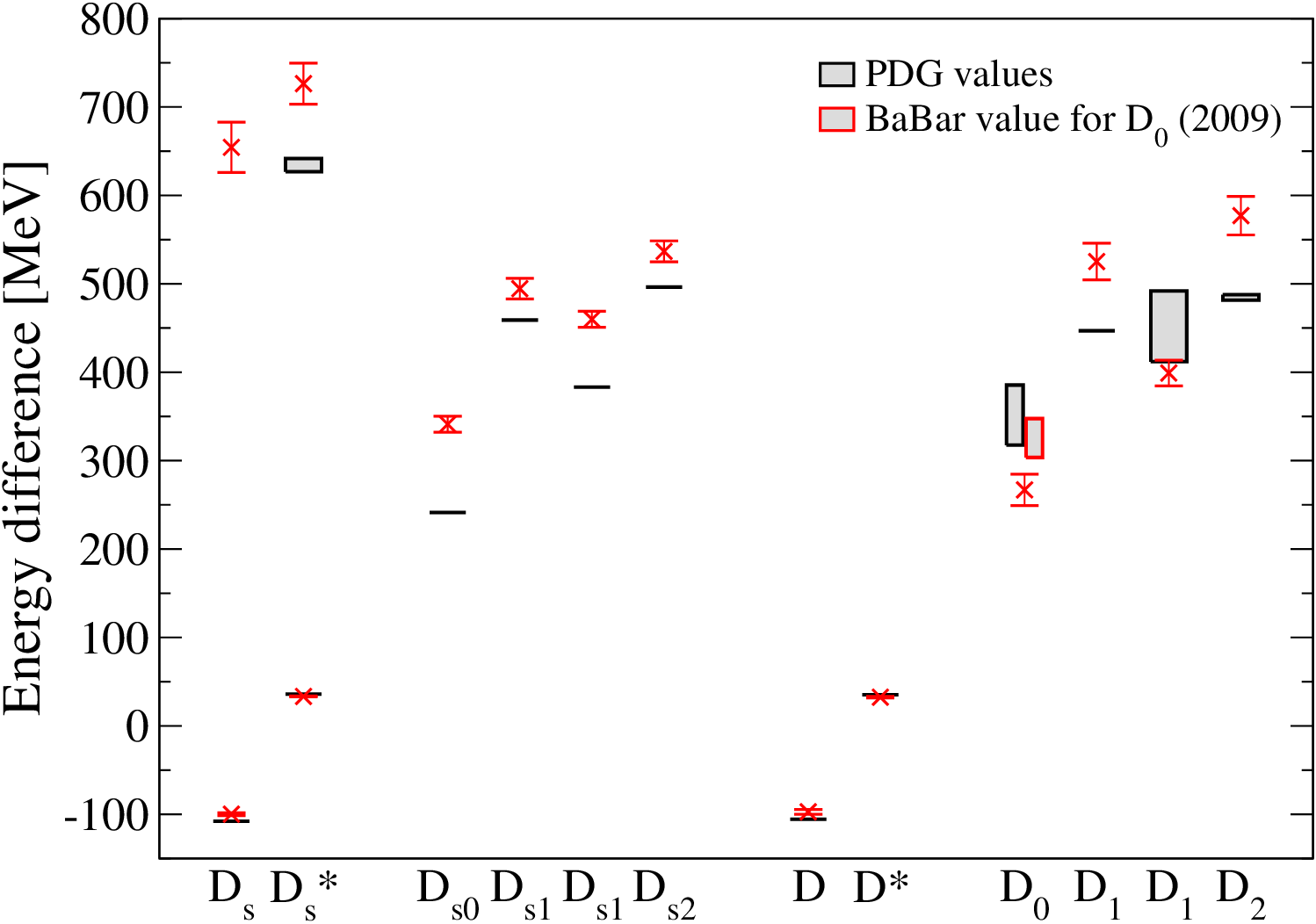}
    \caption{Left panel: Mass splittings in the charmonium spectrum compared to the spin-averaged ground state mass $(M_{\eta_c}+3M_{J\Psi})/4$. Right panel: Same for charmed mesons.}
    \label{overviewplots}
  \end{center}
\end{figure}

\begin{figure}[tbh]
\begin{center}
\includegraphics[width=85mm,clip]{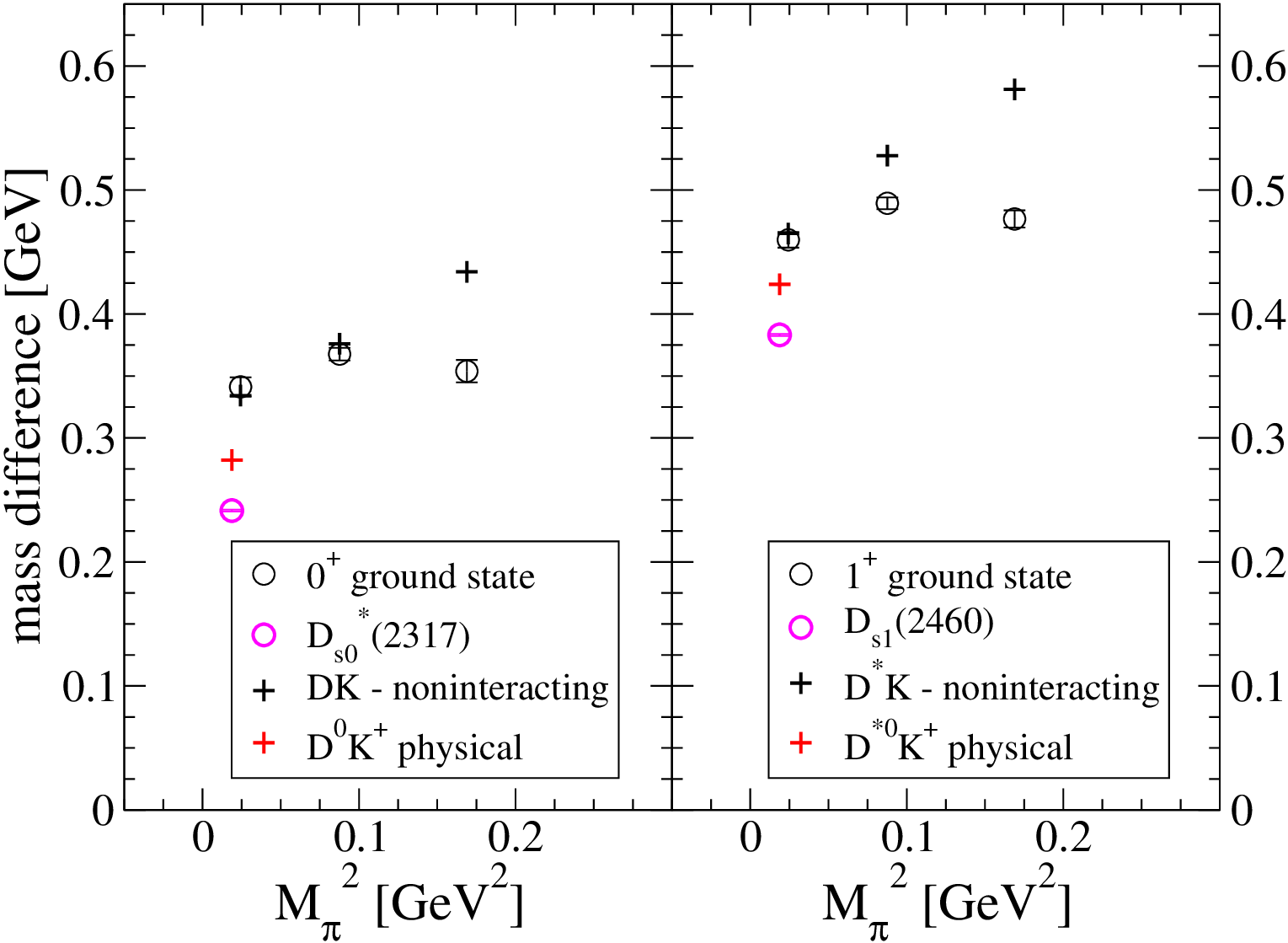}
\caption{Measured energy levels for the $D_{s0}^\star$ (left panel) and $D_{s1}$ (right panel) ground states (black circles) compared to experimental states (magenta circles). All masses are plotted with respect to the spin-averaged $D_s$ ground state. The plus signs denote the $DK$ and $D^\star K$ scattering levels on the lattice (black) and in nature (red). At our lowest pion mass the artificially heavy scattering states are very close to the measured ground state energy.} 
\label{hs_scattering}
\end{center}
\end{figure}

The right panel of Figure \ref{overviewplots} shows the $D$ and $D_s$ spectrum as extracted from our simulation on the ensemble with the lightest sea quark mass. We obtain reasonable values for the hyperfine splittings and for the pairs of states corresponding to the multiplet with $j^P=\frac{3}{2}$ in the heavy quark limit. While light sea quarks are important for several of the states, the doublets corresponding to $j^P=\frac{1}{2}$ in the heavy quark limit show a substantial difference compared to experiment, which is hard to explain by discretization effects alone. For these states the nearby $DK$ and $D^\star K$ thresholds may play an important role. Figure \ref{hs_scattering} compares the measured energy levels in the lattice simulation with the experimental resonance and the relevant scattering threshold from experiment. It can be seen that the scattering threshold in our simulation is slightly unphysical. While the state we observe seems to coincide with the scattering threshold at the smallest pion mass, this is no longer the case for larger pion masses and the overlaps of the state with our interpolator basis (as encoded in the eigenvectors of the GEVP) suggests that we see the same state at all pion masses. In light of this, further studies should include the relevant multi-meson states in the variational basis, which is challenging from a computational point of view.

\acknowledgements{%
We thank the PACS-CS collaboration for access to their gauge configurations. The calculations were performed on computing clusters at TRIUMF and York University. We thank Sonia Bacca and Randy Lewis for making these resources available. This work is supported in part by the Natural Sciences and Engineering Research Council of Canada (NSERC).
}



%

}  
